\documentstyle[twocolumn,aps]{revtex}
\begin{document}
\begin{title}
{\bf Collective Edge Excitations In The Quantum Hall Regime:\\
Edge Helicons  And Landau-level Structure }
\end{title}
\author{O. G. Balev$^{*}$ and P. Vasilopoulos$^\dagger$}
\address{$^{*}$Institute of Physics of Semiconductors, National
Academy of Sciences,
45 Prospekt Nauky, Kiev 252650, Ukraine
\ \\
%\ \\
$\ ^\dagger$Concordia University, Department of Physics, 1455 de
Maisonneuve Blvd O, Montr\'{e}al, Qu\'{e}bec, Canada, H3G 1M8}
%\date{May 21, 1997}
\address{}
\address{\mbox{}}
\address{\parbox{14cm}{\rm \mbox{}\mbox{}\mbox{}
Based on a {\it  microscopic} evaluation of the local current density, 
a treatment of edge magnetoplasmons (EMP)
is presented for %very low temperatures and sufficiently steep 
confining potentials that allow  Landau level (LL) flattening to be neglected.
Mode damping due to electron-phonon interaction is evaluated. 
For $\nu=1, 2$ there exist independent modes spatially
{\it symmetric} or {\it antisymmetric}  with respect to the edge.
Certain modes, changing shape during propagation, 
are  nearly undamped  even for very strong dissipation
 and  are  termed {\bf edge helicons}. 
 For $\nu> 2$ inter-LL Coulomb coupling leads to a strong
 repulsion of the {\bf decoupled} LL fundamental modes.
The theory agrees well with recent experiments.}}
%\end{abstract}
\address{\mbox{}}
\address{\parbox{14cm}{\rm PACS numbers:}}
\maketitle

The essentially classical treatments \cite{1}-\cite{2} of low-frequency
 collective excitations, propagating along the edges of a two-dimensional
electron gas (2DEG) subject to a normal magnetic field $B$, termed  in Ref. \cite{5a} 
edge magnetoplasmons (EMP), account for some important characteristics of
EMP, e.g., the gapless spectrum of these excitations \cite{1} and the
{\it acoustic} modes \cite{2}, \cite{3}.  However, the results of Refs.
\cite{1} and \cite{2} are valid, respectively, for infinitely sharp
and smooth density profiles that are independent of the filling factor
$\nu$.  As contrasted in Fig.  1 with our calculated density profile
for one or two LLs occupied and a smooth ,
 on the magnetic length $\ell_{0}=\sqrt{\hbar /|e|B}$ scale, 
parabolic confining potential %edge, %confining potential, 
these assumed profiles miss an important quantum mechanicall  aspect, 
the LL structure. This inadequacy was manifested in the
observed \cite{3} plateau structure of the transit times reflecting
that of the quantum Hall effect (QHE) plateaus and not accounted for
in Ref.  \cite{2}.  In addition, for a spatially homogeneous
dissipation within the channel, the damping is found quantized and
independent of temperature \cite{1} or it is treated
phenomenologically \cite{2} with damping rates strongly overestimated
\cite{3}.  Other limitations of the model of  Ref. \cite{1}
were pointed out in Refs. \cite{4}-\cite{5}. 
In a sense the conventional EMPs \cite{1}-\cite{5a} is the magnetic analog of the
Kelvin wave  \cite{5c}  at the edge of a rotating
"shallow" sea  with   chirality  determined by the Coriolis parameter 
which corresponds to  
the cyclotron frequency $\omega_{c}=|e|B/m^* $.  In these mostly 
{\it classical} models the position of the edge does not vary but
the charge density profile at the edge does.

 In another distinctly different and fully quantum-mechanical edge
  wave mechanism [8-10] only the edge position, for $\nu=1$, of an
 incompressible 2DEG varies; with respect to that the density profile
 is that of the undisturbed 2DEG. This approach  is limited
to the subspace of the lowest LL wave functions, neglects LL mixing and
dissipation, and results in a single chiral EMP  
with dispersion law similar to that  in \cite{1}. %-\cite{2}.

 Both previous classes of models are oversimplifications. In this Letter
we present a quasi-microscopic treatment of EMPs for integer $\nu$, 
which takes into account LL structure, 
LL mixing,  dissipation (related to  LL mixing essentially),  and 
the inhomogeneity of the current density near the edges treated
recently \cite{7}.  It  is valid for bare confining potentials  
sufficiently steep  that LL flattening  and the formation of compressible
and incompressible strips$^{ \cite{8}}$ can be neglected$^{ 
\cite{9}}$; in this case the dissipation is essential only within a distance
$\alt \ell_{0}$ from the edges \cite{7}.
As will be made clear, our model effectively
incorporates  the previous two distinct propagation mechanisms. 

We consider a zero-thickness 2DEG, of width $W$ and of length $L_x=L$,
in the presence of a strong magnetic field $B$ along the $z$ axis.  We
take the confining potential flat ($V_{y}=0$) in the interior of the 2DEG
 and parabolic at its edges,  $V_{y} =m^*\Omega^2
(y-y_r)^2/2$, $y\geq y_r$.  $V_{y}$ is assumed smooth on the scale of
$\ell_{0}$ such that $\Omega \ll \omega_{c}$. The resulting
one-electron energy spectrum $ E_{n}(k_{x})= (n+1/2)\hbar
\omega_{c}+m^*\Omega^2 (y_{0}-y_{r})^2/2$, where $y_{0}=\ell_{0}^{2}
k_{x} \geq y_{r}$, leads to the group velocity of the edge states
$v_{gn}=\partial E_{n}(k_{e}^{(n)})/\hbar\partial k_{x}=\hbar
\Omega^{2}k_{e}^{(n)}/m^{*}\omega_{c}^{2}$ with characteristic wave
vector $k_{e}^{(n)}=(\omega_{c}/ \hbar\Omega)
\sqrt{2m^{*}\Delta_{Fn}}$, $\Delta_{Fn}=E_{F}-(n+1/2)\hbar
\omega_{c}$. The edge of the n-th LL is denoted by
$y_{rn}=y_{r}+\ell_{0}^{2}k_{e}^{(n)}$ and $W=2y_{r0}$.

Assuming $|q_{x}| W \gg 1$, we can consider an EMP along the right
edge of the channel of the form $A(\omega, q_{x}, y) \exp[-i(\omega
t-q_{x} x)]$, totally independent of the left edge.  We % consider
% at most  two LL   occupied and
neglect the spin-splitting for $\nu$ even.
Because the wavelength $\lambda$
of the practically quasi-static EMP satisfies $\lambda\gg \ell_{0}$,
the electric field $E_{x}(\omega, q_{x},y)$ has a smooth $y$
dependence on the scale of $\ell_{0}$.  %Then
Following Ref. \cite{7}
we obtain the current density $j_\mu$ in the form
$j_y(y)=\sigma_{yy}(y) E_{y}(y)+ \sigma_{yx}^0(y) E_{x}(y)$,
$j_x(y)=\sigma_{xx}(y)E_x(y)- \sigma_{yx}^0(y) E_{y}(y)+
\sum_{j}v_{gj}\delta\rho_{j}(\omega, q_{x},y)$.
The convection contribution
$v_{gj}\delta\rho_{j}$  %to $j_x$
is due to  a charge distortion $\delta\rho_{j}$
 localized near the edge of the $j$-th LL.
These contributions to  $j_\mu$ are microscopically obtained when
$E_{\mu}(y)$   is smooth on the scale of $\ell_{0}$. This  holds
for the components $\propto E_x(y)$   but is not well justified for those
$\propto E_y(y)$. We approximate the latter
by those obtained when $E_y(y)$ is smooth.  
This is equivalent to  neglecting %possible
nonlocal contributions to $j_\mu\ $
$\propto \int dy' \sigma_{\mu y}(y,y') E_{y}(y')$. For weak dissipation  
the  results for the fundamental modes can   
be justified by a {\it microscopic} RPA treament
\cite{10} which includes nonlocal effects
 and does not  require the smoothness of $E_{\mu}(y)$ on the scale of 
$\ell_{0}$. The Hall conductivity %$\sigma_{yx}^0(y)$ is \cite{7}

                \begin{equation}
                \sigma_{yx}^0(y)=\frac{e^2}{2\pi\hbar}
                \sum_{n}\int_{-\infty}^\infty dy_{0}f(E_{nk_{x}})
                \Psi_{n}^2(y-y_{0}),
                \label{5}
                \end{equation}
where $\Psi_{n}(y)$ is a harmonic oscillator function and
$f(E_{nk_{x}}) $ the Fermi-Dirac function.  We consider only the
interaction of electrons with piezoelectric phonons  and neglect that
with impurities shown to be very weak \cite{11}.
We approximate $\sigma_{xx}(y)$
by $\sigma_{yy}(y)=\sum_{n}\sigma_{yy}^{(n)}(y)$ and calculate it
for very low temperatures $T \ll \hbar v_{gn}/\ell_{0} k_{B}$
using Ref.  \cite{7}. For $v_{gn}> s$ and $\nu=2,4$ we obtain
$\sigma_{yy}^{(n)}(y)=\tilde{\sigma}_{yy}^{(n)}\Psi_{n}^2(\bar{y}_{n})$,
$\bar{y}_{n}=y-y_{rn}$, and $\tilde{\sigma}_{yy}^{(n)}= 3
e^{2}\ell_{0}^{4}c'k_{B}^{3}T^{3}/\pi^{2}\hbar^{6} v_{gn}^{4}s$ where
$s$ is the speed of sound  and  $c^{'}$ the %electron-phonon
interaction constant. 

Using $j_{\mu}$, the continuity equation linearized in
$\delta\rho\equiv \rho$, and Poisson's equation we obtain the integral
equation

                \begin{eqnarray}
                \nonumber
                -i\sum_{n}&&(\omega-q_{x}v_{gn})\rho_{n}(\omega,
q_{x},y)+\frac{2}{\epsilon}
                \Big[ q_{x}^{2}\sigma_{xx}(y)\\*
                \nonumber
                &&-iq_{x}\frac{d}{d y}[\sigma_{yx}^{0}(y)]
                -\sigma_{yy}(y)\frac{d^{2}}{d y^{2}}
                -\frac{d}{d y}[\sigma_{yy}(y)]\frac{d}{d y}\Big]\\*
                &&\times\int_{-\infty}^{\infty}
                dy' K_{0}(|q_{x}||y-y'|)\rho(\omega, q_{x},y')=0,
                \label{7}
                \end{eqnarray}
where $\epsilon$ is the spatially homogeneous dielectric constant.
For a dissipationless, classical 2D electron liquid Eq.  (\ref{7})
becomes identical with Eq.  (4) of Ref.  \cite{2}. % Further, i
If $\sigma_{\mu\nu}^{0}(y)$ is independent of $y$, for $|y|<W/2$, Eq.
(\ref{7}) reduces to Eq.  (15) of Ref.  \cite{1}.
To solve Eq. (\ref{7}), we remark that for  $\hbar v_{gn}\gg \ell_{0}
k_{B} T$ we have $d [\sigma_{yx}^{0}(y)]/dy\propto
[\Psi_{0}^{2}(\bar{y}_{0}) +\Psi_{1}^{2}(\bar{y}_{1})]$.  It follows
%from Eq.  (\ref{7})
that $\rho_{n}(\omega,q_{x},y)$ is concentrated
within a region of extent $\sim \ell_{0}$ around the edge of the $n$th LL.  
For $2\Delta y=y_{r0}-y_{r1}\gg\ell_{0}$, cf.  Fig.
1, we   neglect the exponentially small overlap between
$\rho_{0}(\omega,q_{x}, y)$ and $\rho_{1}(\omega,q_{x}, y)$ and,
for $\nu=4$, attempt the exact solution

        \begin{eqnarray}
                \nonumber
                \rho(\omega,q_{x},y)&=&\Psi_{0}^{2}(\bar{y}_{0})
                \sum_{n=0}^{\infty}                 
                \rho_{0}^{(n)}(\omega,q_{x}) H_{n}(\bar{y}_{0}/\ell_{0})\\* 
	&&+\Psi_{1}^{2}(\bar{y}_{1}) \sum_{l=0}^{\infty}
                \rho_{1}^{(l)}(\omega,q_{x}) H_{l}(\bar{y}_{1}/\ell_{0}),
                \label{8}
                \end{eqnarray} 
where $H_{n}(x)$ are the  Hermite polynomials. 
We call the terms $l=0,\ 1, \ 2$, etc., the
monopole, dipole, quadrupole, etc.  terms in this expansion of
$\rho_{n}(\omega,q_{x},y)$. %about $y=y_{rn}$.

We now multiply Eq. (\ref{7})  by $H_{m}(\bar{y}_0/\ell_{0})$
and integrate over $y$.  
This procedure is repeated with $H_{k}(\bar{y}_1/\ell_{0})$. With the
abbreviations  $\rho_{0}^{(m)}(\omega, q_{x})\equiv \rho_{0}^{(m)},\
a_{mk}( q_{x})\equiv a_{mk} $ etc., we obtain,
the  coupled systems of equations
        %\begin{eqnarray}
                \begin{equation} %\nonumber
                %(\omega-q_{x}v_{g0})
                \bar{\omega}_0\rho_{0}^{(m)} %&& -(S_{0}+mS_{0}^{`})
         - S_{0m} \sum_{n=0}^{\infty}c_{mn}\Big[a_{mn} \rho_{0}^{(n)} %\\*
                %&&-(S_{0}+mS_{0}^{`})
                %-S_{0m}
                %\sum_{l=0}^{\infty}
                %+c_{mn}
                +b_{mn} \rho_{1}^{(n)}\Big]
                =0,
                \label{9}
                \end{equation} %\end{eqnarray}
        \begin{eqnarray}
                \nonumber
                && %(\omega-q_{x}v_{g1})
\bar{\omega}_{1} \Big[A_{k} \rho_{1}^{(k)}+B_{k}
                \rho_{1}^{(k +2)}+\rho_{1}^{(k -2)}/2\Big]/2 \\*
                %\nonumber
                 %-(S_{1}+mS_{1}^{`})
             && -\sum_{n=0}^{\infty} c_{kn} \Big[S_{1k}F_{nk}/2
                -\sqrt{k}S_{1}^{`}\tilde{F}_{nk}\Big]=0.
                 \label{10}
                \end{eqnarray}
                %\ b_{nm_{1}}\rho_{0}^{(n)}
                %+\sum_{j=0}^{\infty}c_{m_{1}j}
                %+d_{m_{1}n}\rho_{1}^{(n)}\Big]/2\\*
        %&& +\sqrt{m_{1}}S_{1}^{`} 
        %\sum_{n=0}^{\infty}c_{m_{1}n}\Big[
               %\tilde{b}_{nm_{1}}\rho_{0}^{(n)}
                %+\sum_{j=0}^{\infty} c_{m_{1}j}
                %+\tilde{d}_{m_{1}n}
%\rho_{1}^{(n)}\Big]=0,
                %\label{10}
                %\end{eqnarray}
Here  $\bar{\omega}_n=\omega-q_{x}v_{gn},\ F_{nm}=b_{nm 
}\rho_{0}^{(n)}+d_{mn}\rho_{1}^{(n)}, \ \tilde{F}_{nm}=
\tilde{b}_{nm }\rho_{0}^{(n)} +\tilde{d}_{m n}, \ S_{nm}=S_{n}+mS_{n}^{`}$, 
  $S_{n}=2(q_{x}\sigma_{yx}^{0}-i
 q_{x}^{2}\tilde{\sigma}_{xx}^{(n)})/\epsilon$,
 $S_{n}^{`}=-4i\tilde{\sigma}_{yy}^{(n)}/\epsilon\ell_{0}^{2}$, 
 and $\sigma_{yx}^{0}=e^{2}/\pi\hbar$.
 $a_{mn}$ is given in Ref. \cite{12}  and $b_{mn}$, $\tilde{b}_{mn}$,
$d_{mn}$, and $\tilde{d}_{mn}$  are given by similar expressions.  
 Further, $c_{mn}=(2^{n}n!/2^{m}m!)^{1/2}$, $A_{m }= (2m +1)$,
 and $B_{m}=(m +2)(2m +2)$. 

${\bf (i)\ \nu=2}$.  In this case the second term of Eq.  (\ref{8}),
the third term of Eq.  (\ref{9}), and Eq.  (\ref{10}) are absent.
Eqs.  (\ref{8}), (\ref{9}), and the form of $a_{mn}$ show \cite{12}
that there exist
{\it independent}  modes, spatially {\it symmetric},
$\rho^{s}(\omega,q_{x},y)$, or {\it antisymmetric},
$\rho^{as}(\omega,q_{x},y)$, with respect to $y=y_{r0}$; they correspond to
 $n$ {\it even} or {\it odd}, respectively.

{\it Symmetric modes}.  We first consider only two terms,
$n=0$ and $n=2$, in  Eq.  (\ref{8}).
 For $m=0$ and $m=2$ Eq.  (\ref{9}) gives
a system of two coupled equations for the unknows $\rho_{0}^{(0)} $
and $\rho_{0}^{(2)}$.  The vanishing of the determinant %of the coefficients
gives two branches $\omega_{+}^{s} $ and
$\omega_{-}^{s}$.  With $v_{g}\equiv v_{g0}$ and $S\equiv S_{0}$
their dispersion relations (DRs)
 read \cite{12} $\omega_{\pm}^{s}=q_{x}v_{gn}+\{ R_{+}\pm [R_{-}^{2}
+4S(S +2S^{`})a_{02}^{2}]^{1/2}\}/2$, where $R_{\pm}=[S(a_{00}\pm
a_{22})\pm 2S^{`}a_{22}]$.

{\bf Edge helicons}. The coupling between the branches
(due to $a_{02} \neq 0$) and the
strength of the dissipation  modify  the character of the pure modes.
For $K\gg \eta,\ \eta=\tilde{\sigma}_{yy}^{(0)}/\ell_{0}^{2}\sigma_{yx}^{0}
|q_{x}|$, and {\it weak dissipation} $\eta < 1/4$, the
 $\omega_{-}^{s}$ branch remains almost unchanged whereas the
 $\omega_{+}^{s}$ branch acquires a principally new contribution to damping
 since $\omega_{+}^{s}  =q_{x}v_{g} +S\ (K+1/4) +S^{`}/4K$, 
 $K=1/2-\ln(q_{x} \ell_{0})$.
The coupling  leaves the phase velocity of  both branches
nearly unchanged and the $\omega_{+}^{s}$ branch is {\it very weakly damped}
and almost monopole-like since
$\rho_{0}^{(0)}/\rho_{0}^{(2)}\approx -8\ K$, $K \gg 1$.
For {\it strong dissipation} ($K\gg \eta\gg 1/4 $)
we obtain $\rho_{0}^{(0)}/\rho_{0}^{(2)}\approx -2\ i K/\eta$.
This  corresponds to  $\omega_{+}^{s}\tau_{0}^{*}\gg \nu
r_{0}/\pi\gg \omega_{+}^{s}\tau_{0}^{*}/(4K+1)$, where
$r_{0}=e^{2}/\epsilon\hbar\omega_{c}\ell_{0}$, and
$\omega_{+}^{s}$ can still be considered   high compared to
$1/\tau_{0}^{*}$; $\tau_{n}^{*}$, defined by
$1/\tau_{n}^{*}=\omega_{c}\tilde{\sigma}_{yy}^{(n)}/(\sigma_{yx}^{(0)}
\ell_{0}\sqrt{2n+1})$, is an effective scattering
time in an edge strip of width $\ell_{0}\sqrt{2n+1}$. In this frequency region 
we call the $\omega_{+}^{s}\equiv \omega_{EH}^{(0)}$
branch  high-frequency {\it edge helicon} (HFEH). Due to the almost $\pi/2$
shift between $\rho_{0}^{(0)}$ and $\rho_{0}^{(2)}$, the HFEH exhibits the
following remarkable property: if its charge  along $y$ has
a pure quadrupole character $\propto |\rho_{0}^{(2)}|$ for some phase of
the wave, after approximately a $\pm\pi/2$ shift it acquires a pure
monopole character $\propto |\rho^{(0)}|$. Notice that 
Im$\omega_{EH}^{(0)}\propto T^{3}$.  
That is, in contrast with Ref.  \cite{1}, the damping of the HFEH
scales with $T$ and is not quantized in the QHE plateaus.
As for the $\omega_{-}^{s}$ branch, it is strongly damped.

For {\it very strong dissipation},
$\eta\gg K$, the $\omega_{-}^{s} $ branch is strongly damped while
 the $\omega_{+}^{s}$ branch changes to  
a low-frequency {\it edge helicon} (LFEH) with DR
($\omega_{+}^{s}\equiv\omega_{EH}^{LF}$) 

        \begin{equation}
        \omega_{EH}^{LF}=q_xv_g+[S%%%(K-1/4) %\frac{1}{4})
                -i\tilde{\sigma}_{yy}^{(0)}/\eta^{2}\ell_{0}^{2}\epsilon]
(K-1/4) ,  
                \label{13}
                \end{equation}
  where $\omega_{EH}^{LF}\tau_{0}^{*}\ll \nu r_{0}/\pi\alt 1$. Despite 
 this, the LFEH is {\it very weakly damped}.  
 Further,  $\rho_{0}^{(0)}/\rho_{0}^{(2)}\approx 2$ and
 Eq. (\ref{8}) gives the charge density profile $\delta \rho=
\sqrt{\pi}\ell_{0}\ $Re$[\rho(\omega,q_{x},y)/\rho_{0}^{(0)}(\omega,q_{x})]$  
shown in Fig. 2 by curve 2  
for $K/\eta=0.01$; such a small ratio has practically no effect
on $\delta \rho(y)$ if only the terms $n=0$ and $n=2$ are kept in Eq. (\ref{8}).
Since $\delta \rho(y)$   is symmetric with respect to the edge, only
one half of Fig. 2 is shown. Curve 1 shows the
monopole term ($\ \propto \Psi_{0}^{2}$ ). 
The effective %for all essential regios of
convergence parameter for curve 2 is not sufficiently small. 
To better describe the profile of the LFEH
 we also plot curves 3, 4, and 5   obtained
with 3, 4, and 5 even $n$ terms  
retained in Eq. (\ref{8}), respectively. 
As shown, keeping 4 or 5 terms in the n-summation leads
already to a clear convergence in the form of the charge-density
profile, without altering its oscillatory character or
 changing its magnitude by much.  This  oscillatory behavior of
$\delta \rho$, further modified during propagation, is in sharp contrast with 
the ``usual'' EMPs of Ref.  \cite{1} and the $j=0$ mode of Ref.  \cite{2}.
Equation (\ref{13}) already   approximates well
Re$\omega_{EH}^{LF}$ and   the dependence of Im$\omega_{EH}^{LF}$ on $T$.

The {\it antisymmetric} modes have been described in Ref. \cite{12}.  
Here it is worth mentioning that 
if we keep only one, two, or three {\it odd} terms in Eq. (3), the
dimensionless velocity of the dipole branch $v_{dip}=
(\omega/q_{x}-v_{g0})/(e^{2}/\pi \hbar \epsilon)$
for weak dissipation is equal, respectively, 
to $0.4996$, $0.5963$, and  $0.6287$; the charge density profile
shows a similar fast convergence.

It is worth noticing  that  if we  limit ourselves  to the subspace 
of the $n=0$ LL wave functions,  by keeping, for $\nu=1$, only the $n=0$ 
 term in Eq. (\ref{8}), we  have the same edge-wave mechanism as
Refs. \cite{6}-\cite{6b} with the same  single mode.
This can be seen by writing 
$n(x,y,t)=n_0(y+b(\omega,q_x) \cos(\omega t- q_x x))
\approx n_0(y) +[dn_0(y)/dy] b(\omega,q_x) \cos(\omega t- q_x x)$,
with $dn_0(y)/dy\sim \psi_0^2(\bar{y}_0)$, for the total density $n(x,y,t)$. 
It is only by retaining the $n \geq 1$ terms that we  
obtain  more than one modes with important  contributions to the 
damping  of the  fundamental mode. Further, retaining  the $n \geq 1$ terms 
is equivalent to incorporating in the model the classical
edge-wave mechanism \cite{1}-\cite{5a}, \cite{5c}.
It is also clear that we focus on wave effects of non-spin nature and
do not treat spin excitations such as skyrmions.

{\bf (ii) inter-LL coupling: $\nu=4$}.  Although the condition $2\Delta
y\gg\ell_{0}$, cf.  Fig.  1, is well justified for $V_{y}$, 
the system of Eqs.  (\ref{9}) and (\ref{10})
can be strongly coupled due to the long-range nature of the Coulomb
interaction.  To make contact with the ${\bf \nu=2}$ results, we first
 consider the {\it symmetric} modes, $\rho_{1}^{(0)}$ and
$\rho_{1}^{(2)}$ of the $n=1$ LL {\bf decoupled} from the $n=0$ LL.
Then one  branch  is $\omega_{3}^{(1)}\approx
q_{x}v_{g1} +(S_{1}+2S_{1}^{`})/4$.
 The other one is the fundamental branch, or HFEH of the $n=1$ LL,
 $\omega^{(1)}_{EH}\approx q_{x}v_{g1}
+S_{1}(K-1/4) +S_{1}^{`}/12K$.
Now the {\bf decoupled} fundamental modes 
of $n=0$ and $n=1$ LLs have DRs given
by $\omega^{(0)}_{EH}$ and $\omega^{(1)}_{EH}$.  
When they are {\bf coupled}, %by the inter-LL mode coupling,
their DRs change drastically.
For $2\Delta y\gg \ell_{0}$ and $2\Delta y \; q_{x} \ll 1$
an examination of the coefficients $a_{mn}$ etc. 
 shows that the most important terms in  
Eq.  (\ref{8}) are $\rho_{0}^{(0)}$, %$\rho_{0}^{(2)}$,
$\rho_{1}^{(0)}$, and $\rho_{1}^{(2)}$. 
This leads to %three coupled equations giving
three branches, $\tilde{\omega}^{(01)}_{\pm}$ and
$\omega^{(01)}_{3}\approx \omega^{(1)}_{3}$.
The renormalized $n=0$ LL fundamental mode becomes
$\tilde{\omega}^{(01)}_{+}\approx q_{x} (v_{g0}+v_{g1})/2+
(2/\epsilon) q_{x}\sigma_{yx}^{0}[2\ln(1/q_{x}\ell_{0})-
\ln(2\Delta y/\ell_{0})+3/5]+S_{1}^{`}/16K$  and
that of the $n=1$ LL $\omega^{(01)}_{-}\approx q_{x}
(v_{g0}+v_{g1})/2+ (2/\epsilon) q_{x} \sigma_{yx}^{0}[\ln(2\Delta
y/\ell_{0})+2/5]+
S_{1}^{`}/\{24[\ln(\Delta y/\ell_{0})+\gamma+1/4]\}$, where 
$\gamma$ is the Euler constant. The $\omega^{(01)}_{-}$ mode
becomes purely acoustic and has a phase velocity larger than that of
the $j=1$ mode of Ref.  \cite{2} for $2\Delta y/\ell_{0}\geq 5$.
The coupled fundamental modes $\tilde{\omega}^{(01)}_{\pm}$
are very weakly damped. 

The DRs for $\nu=4$, corresponding to the experimental \cite{3}
parameters $B=2.06$ Tesla and $T=1.5 K$, are shown in Fig.  3.  The
solid and short-dashed curves are obtained with $\epsilon=12.5$.  The
dashed curves ($\epsilon=6.75$) pertain to a sample with air above the
spacer.  The short-dashed curves are the {\bf decoupled} fundamental
modes, the solid and dashed ones the {\bf coupled} modes.  As can be
seen, the inter-LL coupling   strongly modifies the DR of both fundamental
modes.  Using  $\Omega=7.8\times 10^{11}$/sec \cite{17} gives
$\Omega/\omega_{c}\approx 0.14$, $2\Delta y/\ell_{0}\approx 6$,
$v_{g0}=2.3 \times 10^{6}$/sec, and $v_{g0}/v_{g1}= \sqrt{3}$.  The
 $\nu=4$ modes,  in Fig.  (3a) of Ref.  \cite{3}, are very
well described by the renormalized  fundamental modes
$\tilde{\omega}_{\pm}^{(01)}$.
The same holds for the  $\nu=4$  modes of Fig.  3 (b) of Ref.  \cite{3}.
The mode $\omega_{3}^{(01)}$ is strongly damped: with
$\epsilon=6.75$ its decay rate is  Im$ S_{1}^{`}/2\approx
2\tilde{\sigma}_{yy}^{(1)}/\epsilon\ell_{0}^{2}\approx 1.3\times
10^{10}$/sec.  This is smaller than that of the $j=1$ branch of Ref.
\cite{2} $1/\tau_{1}\approx 2\times 10^{10}$/sec.   
The decay rate of the $j=0$ mode is
$1/\tau_{0}\approx 1.7\times 10^{9}$/sec whereas those of the
$\tilde{\omega}_{\pm}^{(01)}$ modes are about ten times smaller,
Im$\tilde{\omega}_{+}^{(01)}\approx 2.1\times 10^{8}$/sec and
 Im$ \tilde{\omega}_{-}^{(01)}\approx 5.6 \times 10^{8}$/sec
$\ll 1/\tau_{1}\approx 2 \times 10^{10}$/sec.
Thus, the  decay rates of the $\tilde{\omega}_{\pm}^{(01)}$ modes
should be much closer to those of the experiment  \cite{3}
than the strongly overestimated  ones \cite{2}.
  Regarding the delay times $t_{d}$ for the
sample with length $L_{x}=320 \mu$m, we obtain $t_{d}=1.2\times
10^{-10}$sec for the $\tilde{\omega}_{+}^{(01)}$ mode and $t_{d}=6.9\times
 10^{-10}$sec for the
$\tilde{\omega}_{-}^{(01)}$ mode, in very good agreement with the
observations \cite{3}.  We conclude that the slower mode observed for
$\nu=4$ is not the $j=1$ mode of Ref.  \cite{2} but the present
$\tilde{\omega}_{-}^{(01)}$ mode. It is also clear that our theory accounts
for the  existence of the plateaus in $t_{d}$ \cite{3}  as the
quantized Hall conductivity appears  in all DRs.

In summary, we  presented a  theory of edge
magnetoplasmons for  confining potentials that allow  
LL flattening to be neglected.
It accounts for the existence of plateaus
in the delay times, the dispersion relations, and the damping rates
of the observed \cite{3} modes for $\nu=4$.
Compared to the {\it decoupled}, individual LL fundamental modes,
the {\bf coupled}  LL modes are drastically renormalized and
in good agreement with the experiment. Other novel results are mentioned
in the abstract. 

This work was supported by NSERC Grant No. OGP0121756. 
O G B acknowledges partial support from the Ukrainian SFFI Grant No. 2/4/665.

\begin{figure}
\caption{Unperturbed electron density $n_{0}(y)$, normalized to the
bulk value $n_{0}$, as a function of $y/\ell_{0}$. The thick solid curve is
the model of Ref. \protect\cite{1} and the short-dashed curve that of Ref.
\protect\cite{2} %($n_{0}(y)/n_{0}= (2/\pi) arctan[(y_{re}-y)/a]^{1/2}$,
($a/\ell_{0}=20$). The dashed and solid curves show the calculated profile for
$\nu=1, 2$ and for $\nu=4$, respectively. The solid and open dots mark the
edges of the $n=1$ and $n=0$ LLs.}
\label{fig.1}
\ \\
\caption{Dimensionless charge density profile $\tilde{\rho}(y)$
of the low-frequency edge helicon  as a function of $\bar{y}_{0}/\ell_{0}$
for $\nu=2$.
The number of {\it even} terms retained in %the first sum of 
Eq. (3)
is shown next to the curves.} % The  monopole  contribution is curve 1.}
\label{fig.2}
\ \\
\caption{EMP dispersion relations
  pertinent to Ref. \protect\cite{3}  for $\nu=4$. The short-dashed curves
are the
  {\bf decoupled}
  fundamental modes  ( $\epsilon=12.5$). The upper two solid ($\epsilon=12.5$)
  and dashed
  ( $\epsilon=6.75$) curves are the {\bf coupled} fundamental modes. The
  lowest
  solid (dashed) curve is the third branch $\omega^{(01)}_{3}\approx
\omega^{(1)}_{3}$.
  The accessible \protect\cite{3}
  frequencies are below $\omega=0.01 \omega_{c}$.}
\label{fig.3}
\end{figure}
\end{document}